\pdfoutput=1
\documentclass[11pt]{article}

\usepackage[preprint]{acl}

\usepackage{times}
\usepackage{latexsym}
\usepackage[T1]{fontenc}
\usepackage[utf8]{inputenc}
\usepackage{microtype}
\usepackage{inconsolata}
\usepackage{graphicx}

\usepackage{booktabs}

\usepackage{amsmath}
\usepackage{amssymb}
\usepackage{amsthm}
\usepackage{enumitem}
\usepackage{tikz}
\theoremstyle{definition}
\newtheorem{definition}{Definition}
\usepackage{xcolor}
\usepackage{listings}
\usepackage{needspace}
\usepackage{xurl}
\usepackage{multirow}
\usepackage{array}

\newif\ifshowappendix
\showappendixfalse
\newcommand{\withappendix}[1]{\ifshowappendix#1\else\unskip\fi}

\input{macro}

\title{Eliciting and Mitigating Specification Satisficing in Coding Agents}

\iflatexml\let\And\and\fi
\author{Ruixin Wang \\
  Purdue University \\
  \texttt{ruixinw@purdue.edu} \\\And
  Tianyi Zhang \\
  Purdue University \\
  \texttt{tianyi@purdue.edu} \\}

\begin{document}
\maketitle

\begin{abstract}
Coding agents often fail to meet all requirements in the task description.
In some of these failures, the agent considers an implementation choice that meets a requirement, but later decides to commit to another one that does not meet the requirement for a reason it finds acceptable.
We call this \emph{specification satisficing}.
It is challenging to identify and mitigate specification satisficing because the agent accepts such choices for its own reasons, and existing self-correction methods are biased toward them.
We propose \satisfix{}, which guides a coding agent to report these choices and their reasons, and then check each choice against the task description.
We evaluate \satisfix{} on DeepSWE and SWE-bench Pro with five LLMs as backbones and two different agent frameworks, and compare it with three self-correction methods.
\satisfix{} improves the pass@1 of the base agent by an average of 5.09\% on DeepSWE and 7.06\% on the hard subset of SWE-bench Pro.
On DeepSWE, it outperforms the best baseline by 3.98\%.
Our ablation study shows that both steps of \satisfix{} contribute to the improvement.

\end{abstract}

\section{Introduction}

Coding agents \citep{yang2024sweagent,zhang2024autocoderover,wang2025openhands,ruan2025specrover} often face complex tasks in large code repositories.
Such a task can contain many requirements and take hundreds of lines of code spanning multiple files \citep{deng2025swebenchpro,huang2026deepswe}.
Recent studies show that coding agents frequently fail to meet all of these requirements \citep{zeng2025e2edevbench,wang2026solvedissues,huang2026deepswe}.

We observe that coding agents often consider an implementation choice that meets a requirement, but later decide to commit to another one that does not meet the requirement for a reason they find acceptable.
We call this \emph{specification satisficing}.

Self-correction is commonly used to identify and mitigate an agent's mistakes \citep{madaan2023selfrefine,palmeira-ferraz-etal-2024-llm,song-etal-2025-progco}.
Recent studies show that its effectiveness is limited \citep{huang2024cannot,stechly2025self,kamoi2024when,olausson2024selfrepair}.
One reason is that an agent is biased toward its own output when asked to identify its mistakes \citep{xu-etal-2024-pride,panickssery2024llm}.
In specification satisficing, the agent accepts its choice for its own reasons, so it does not regard the choice as a mistake \citetext{\citealp{kaddour2026agentic}; cf.\ \citealp{kunda1990case,pronin2002biasblindspot,ehrlinger2005peering}}.

We propose \satisfix{}, the first approach that helps a coding agent surface and mitigate specification satisficing.
It takes two steps before the agent writes code (Figure~\ref{fig:example}).
First, \satisfix{} asks the agent to report where it concluded that a requirement could be handled differently from what the task states.
For each case, it asks for the requirement, the agent's implementation choice, and the reason for this choice.
Unlike existing self-correction methods, it does not ask the agent to judge its choices.
It puts the agent in the role of an advisor who helps revise the task description, so the agent can report a choice without regarding it as a mistake.
Second, \satisfix{} asks the agent to double-check each reported choice against the task description.
It then asks the agent to decide whether to keep or change each choice, and proceed to complete the task.

We evaluate \satisfix{} on DeepSWE \citep{huang2026deepswe} and SWE-bench Pro v2 \citep{deng2025swebenchpro,scale2026sweprov2}.
We use five LLMs as backbones and two agent frameworks, and compare \satisfix{} with three self-correction methods.
These methods are Self-Refine \citep{madaan2023selfrefine}, DeCRIM \citep{palmeira-ferraz-etal-2024-llm}, and ProgCo \citep{song-etal-2025-progco}.

\satisfix{} improves the pass@1 of the base agent by an average of 5.09\% on DeepSWE and 7.06\% on the hard subset of SWE-bench Pro.
On DeepSWE, it outperforms the best baseline by 3.98\%.
Our ablation study shows that both steps contribute to the improvement.
The pass rate drops by 6.2\% to the level of the base agent when \satisfix{} does not put the agent in the role of an advisor.

\section{Problem Definition}
\label{sec:problem}

\begin{figure*}[t]
\centering
\includegraphics[width=\textwidth]{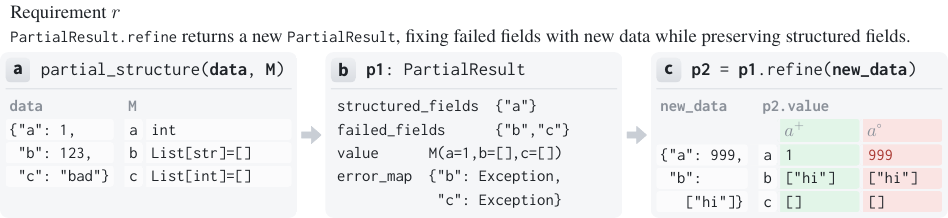}\par\medskip
\includegraphics[width=\textwidth]{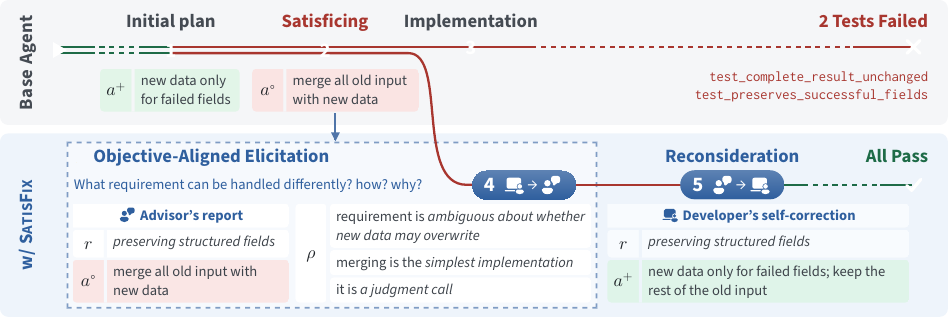}
\caption{Illustration of \satisfix{} with an example task. The task: (a)~partial structuring of \field{data} into class \field{M}, where fields \field{b} and \field{c} fail, (b)~the resulting \field{PartialResult}, and (c)~\field{refine} with new data that fixes \field{b} and also provides the structured field \field{a}. The agent first considers a design $a^{+}$ that preserves structured fields and meets $r$ \circnum{1}, but later commits to a design $a^{\circ}$ that can overwrite them and does not meet $r$ \circnum{2}. Without \satisfix{}, the agent implements this design \circnum{3} and fails two of the 69 evaluation tests. With \satisfix{}, before the agent writes code, it reports the choice in the role of an advisor \circnum{4} and reconsiders it \circnum{5}; its final code passes all 69 tests. The notation follows Section~\ref{sec:problem}.}
\label{fig:task}\label{fig:example}
\end{figure*}

In specification satisficing, a coding agent considers an implementation choice that meets a requirement.
It later commits to another one that does not meet the requirement, for a reason it finds acceptable.
We formalize this definition below.

A task description $D$ states a requirement $r$.
An implementation choice for $r$ is how the agent plans to make the code behave where $r$ applies.
An implementation choice $a$ meets $r$ under premises $P$ if its behavior satisfies $r$ given $P$.
This definition follows the satisfaction argument of \citet{zave1997four}.
The premises $P$ can be the statements in $D$.
They can also be the agent's own premises, and these can include conclusions that $D$ does not state.

The agent considers an implementation choice when it raises the choice as an option.
It commits to the choice when it decides to take it, in its reasoning or in its code.
Let $t$ be the point at which the agent commits to an implementation choice for $r$.
At $t$, the agent holds its own premises $B_t$: what it has taken from $D$, and the conclusions it has drawn while working on the task.
The agent accepts an implementation choice when the choice meets $r$ under $B_t$.

\begin{definition}[Specification satisficing]
\label{def:satisficing}
The agent exhibits specification satisficing on $r$ when three conditions hold.
\begin{enumerate}[nosep,leftmargin=*,label=(\roman*)]
\item\label{cond:i} The agent considers some $a^{+}$ that meets $r$ under $D$.
\item\label{cond:ii} At $t$, the agent commits to another considered choice $a^{\circ}$ that does not meet $r$ under $D$.
\item\label{cond:iii} The agent accepts $a^{\circ}$.
\end{enumerate}
\end{definition}
Condition~\ref{cond:i} requires only that the agent considers $a^{+}$, not that it accepts $a^{+}$.
The term \emph{satisficing} comes from \citet{simon1955behavioral,simon1956rational}, where a decision maker takes an alternative that it finds good enough.
Here the agent accepts $a^{\circ}$ as good enough.
Since $a^{\circ}$ does not meet $r$ under $D$, its acceptance rests on premises in $B_t$ that do not follow from $D$.
We call these premises the \emph{reason} $\rho$.
Table~\ref{tab:reasons} lists four types of reasons.

\paragraph{Example.}
We illustrate the definition with a DeepSWE task \citep{huang2026deepswe}, shown in Figure~\ref{fig:task}.
The coding agent is mini-swe-agent \citep{yang2024sweagent} with DeepSeek V4 Flash \citep{deepseekai2026deepseekv4}.
The task\footnote{\taskid{cattrs-\allowbreak partial-\allowbreak structuring-\allowbreak recovery}} asks for partial structuring in cattrs \citep{cattrs2026}, a Python library that converts dictionaries into objects of typed classes.
When some fields of the input fail to convert, partial structuring still structures the other fields (\panelref{fig:task}{a}).
It returns a \field{PartialResult} that records the structured fields and the failed fields (\panelref{fig:task}{b}).
Here $r$ is a requirement on \field{refine}, a method of \field{PartialResult} (\panelref{fig:task}{c}).
It reads: ``\field{PartialResult.\allowbreak refine(data)} returns a new \field{PartialResult}, fixing failed fields with new data while preserving structured fields.''

The agent first considers $a^{+}$, applying new data only to failed fields (\exref{1}).
It meets $r$ under $D$ (condition~\ref{cond:i}).
With $a^{+}$, a failed field that the new data does not provide loses its original error.
The agent wants to keep this error, a preference that the task does not state.
It turns to another choice $a^{\circ}$, which merges the whole original input with the new data (\exref{2}).
The new data can then overwrite a structured field (\panelref{fig:task}{c}), so $a^{\circ}$ does not meet $r$ under $D$.

When it turns to $a^{\circ}$, the agent writes in its reasoning that $a^{\circ}$ is ``\textit{arguably not `preserving', but likely acceptable}''.
It continues: ``\textit{The phrase `preserving structured fields' I'll interpret as `the existing structured fields remain in the result} \ldots\textit{'}''.
The agent commits to $a^{\circ}$ before it writes code (condition~\ref{cond:ii}).
This interpretation is then a premise in $B_t$.
With it, $a^{\circ}$ meets $r$ under $B_t$, so the agent accepts $a^{\circ}$ (condition~\ref{cond:iii}).
The premise does not follow from $D$, so it is part of the reason $\rho$, an intended reinterpretation (Table~\ref{tab:reasons}).

The agent implements $a^{\circ}$ (\exref{3}) and fails two of the 69 evaluation tests, so the task counts as failed.
\withappendix{Appendix~\ref{app:cases} gives examples of the other three types of reasons.}

\paragraph{A limitation of existing self-correction methods.}
By condition~\ref{cond:iii}, the coding agent accepts $a^{\circ}$, so it does not regard $a^{\circ}$ as a mistake.
Existing self-correction methods ask the agent to review its own work while it still holds $\rho$.
It can accept $a^{\circ}$ again \citep{huang2024cannot}.
To find that $a^{\circ}$ does not meet $r$ under $D$, the agent has to check $a^{\circ}$ against $D$ without relying on $\rho$.

\begin{table}[t]
\centering
\small
\begin{tabular}{@{}>{\raggedright\arraybackslash}p{0.26\linewidth}@{\hspace{0.03\linewidth}}>{\raggedright\arraybackslash}p{0.71\linewidth}@{}}
\toprule
\textbf{Type} & \textbf{Reason $\rho$} \\
\midrule
Intended\newline reinterpretation & The requirement can be interpreted another way, so the implementation still meets it. \\
\addlinespace[3pt]
Narrowed\newline scope & Some scenarios the requirement covers will not arise, so they need not be handled. \\
\addlinespace[3pt]
Lowered\newline standard & A weaker version of the requirement is good enough. \\
\addlinespace[3pt]
Perceived\newline conflict & The requirement conflicts with another one, so it can be weakened in favor of the other. \\
\bottomrule
\end{tabular}
\caption{Four types of the reason $\rho$ in Definition~\ref{def:satisficing}.}
\label{tab:reasons}
\end{table}

\section{\satisfix{}}
\label{sec:method}

\satisfix{} has two steps.
First, \satisfix{} asks the coding agent to report where it concluded that a requirement could be handled differently from what the task states (\S\ref{sec:elicitation}).
Then \satisfix{} asks the agent to check every reported implementation choice against the task description and to complete the task (\S\ref{sec:reconsideration}).
Figure~\ref{fig:example} shows both steps on the example task\withappendix{, and Appendix~\ref{app:prompts} gives the prompts}.

The agent decides how to handle each requirement while it explores and plans.
Once it writes code for an implementation choice, an early mistake carries over into the later steps \citep{zhang2024snowball,mehta2026commit}.
\satisfix{} steps in just before the agent writes code (\exref{3}).
By then, the agent has made many of its implementation choices but has not yet built on them.
\withappendix{Appendix~\ref{app:trigger} gives the detailed procedure.}

\subsection{Objective-Aligned Elicitation}
\label{sec:elicitation}

\satisfix{} sends the coding agent the elicitation message $p_{\text{elicit}}$.
The message asks where the agent concluded that a requirement could be handled differently from what the task states.
For each such place, it asks for the requirement, the agent's implementation choice, and the reason for this choice.
The agent replies with a report
\[E = \mathcal{M}(h \,\|\, p_{\text{elicit}}),\]
where $h$ is the agent's history so far, $\mathcal{M}$ is its LLM backbone, and $\|$ denotes concatenation.
Each item of $E$ reports one such place.

In a case of specification satisficing, the agent accepts $a^{\circ}$ for the reason $\rho$, so in its view $a^{\circ}$ meets $r$.
It works as a developer, and its objective is to complete the task.
For this objective, it does not have anything to report about $a^{\circ}$.
The elicitation message says that it comes from the task's maintainers, who wrote the task description and may revise it.
The message puts the agent in the role of an advisor to the maintainers.
The advisor's objective is to show the maintainers where the task description would need to change.
Reporting $a^{\circ}$ and its reason $\rho$ serves this objective.
People report different information depending on the role they take \citep{anderson1978recall}.
The same holds for LLMs \citep{shanahan2023roleplay,marks2025auditing,cywinski2025secret}.
We call this step objective-aligned elicitation.

Unlike existing self-correction methods, the message does not ask the agent to judge its choices.
The agent can report $r$ and $a^{\circ}$ without regarding $a^{\circ}$ as a mistake.
The reason it gives in this item of $E$ can bring out $\rho$.

In the example of Figure~\ref{fig:example}, the agent reports its choice for \field{refine} (\exref{4}).
It reports that the choice ``\textit{violates a strict reading of `preserve structured fields'}''.
As the reason, it reports that the requirement is ``\textit{ambiguous about whether new data may overwrite an already-structured field}''.
This reason brings out the interpretation in Section~\ref{sec:problem}.
There, the agent reads the requirement as allowing new data to overwrite a structured field.
The agent adds that merging is the ``\textit{simplest implementation}'' and that the choice is ``\textit{a judgment call}''.
These premises do not follow from $D$ and are part of~$\rho$.

\subsection{Reconsideration}
\label{sec:reconsideration}

A reported choice can be a case of specification satisficing.
In that case, the coding agent has already accepted the choice.
The agent uses its own premises to judge whether the choice is a mistake, so it can accept the choice again (Section~\ref{sec:problem}).
As in the first step, \satisfix{} does not ask the agent to figure out which of the reported choices are mistakes.
Instead, it asks the agent to reopen every reported choice and to check it against $D$ before deciding to keep or change it.

The elicitation message in Section~\ref{sec:elicitation} says that the maintainers may revise $D$.
\satisfix{} now tells the agent that they will not.
It puts the agent back in the developer's role, and the agent's code has to meet the task requirements in $D$.

\satisfix{} then sends two messages.
The first message $p_{\text{recon}}(D)$ asks the agent to list the alternatives to each reported choice and the plan to check it against $D$ and the repository.
In a case of specification satisficing, $a^{+}$ can be among the alternatives, since the agent has considered it.
The message asks the agent not to settle any choice in its reply
\[Q = \mathcal{M}(h \,\|\, p_{\text{elicit}} \,\|\, E \,\|\, p_{\text{recon}}(D)).\]
The second message asks the agent to carry out the plan in $Q$, settle each choice, and proceed to complete the task.
\satisfix{} does not send any further messages.
If the agent reports nothing in $E$, \satisfix{} only puts it back in the developer's role and asks it to complete the task\withappendix{ (Appendix~\ref{app:trigger})}.

In the example of Figure~\ref{fig:example}, the agent reads the requirement again (\exref{5}).
It writes in its reasoning: ``\textit{`while preserving structured fields' = don't disturb fields in} \field{structured\_fields}''.
It concludes that new data should be applied only to failed fields.
This choice is $a^{+}$.
Its corrected code keeps the whole original input and applies new data only to failed fields (Listing~\ref{lst:refine}).
The code meets $r$ under $D$ and still keeps the original error of \field{c}.
It passes all 69 evaluation tests.

\begin{lstlisting}[style=py, float=tb, escapeinside={(*}{*)}, caption={\field{refine} in the agent's code before (red) and after (green) self-correction with \satisfix{}.}, label=lst:refine]
def refine(self, data):
(*\diffdel*)-   merged = {**self._data, **data}      (*\color{codecomment}\#~$a^{\circ}$*)
(*\diffadd*)+   key_to_field = self._key_to_field()  (*\color{codecomment}\#~$a^{+}$*)
(*\diffadd*)+   merged = dict(self._data)
(*\diffadd*)+   for key, value in data.items():
(*\diffadd*)+       fname = key_to_field.get(key)
(*\diffadd*)+       if (fname is None
(*\diffadd*)+           or fname in self.failed_fields):
(*\diffadd*)+           merged[key] = value
    return self._converter.partial_structure(
        merged, self._cl)
\end{lstlisting}

\section{Evaluation}
\label{sec:evaluation}

\subsection{Setup}
\label{sec:setup}

\paragraph{Benchmarks.}
We evaluate \satisfix{} on DeepSWE v1.1 \citep{huang2026deepswe}.
It has 113 long-horizon software engineering tasks.
Several popular evaluation platforms use it \citep{artificialanalysis2026coding,epoch2026deepswe,mercor2026deepswe}.
To evaluate the generalizability of \satisfix{}, we also use SWE-bench Pro v2 \citep{deng2025swebenchpro,scale2026sweprov2}.
Its 642 tasks are complex enterprise-level problems.
This version corrects the task descriptions that contradicted their tests \citep{openai2026signalnoise,scale2026sweprov2}.
We report results on the full set and on its official hard subset of 51 tasks.

\paragraph{Models.}
We use five LLMs as backbones.
GPT-5.6-Luna \citep{openai2026gpt56} runs with high reasoning effort and DeepSeek V4 Flash \citep{deepseekai2026deepseekv4} with max reasoning effort.
The other three are Qwen3.8-27B \citep{qwen2026qwen38}, Gemma 4 26B A4B \citep{gemmateam2026gemma4}, and Muse Glimmer 30B \citep{museglimmerteam2026museglimmer30b}.

\paragraph{Baselines.}
We compare \satisfix{} with three self-correction methods.
Self-Refine \citep{madaan2023selfrefine} is the most widely used one that does not need another agent or external feedback.
It asks the model to give feedback on its own output and to revise the output with this feedback over several rounds.
DeCRIM \citep{palmeira-ferraz-etal-2024-llm} decomposes the task description into a list of requirements.
It asks the model to check its output against each requirement and to revise the output where a requirement is not met.
ProgCo \citep{song-etal-2025-progco} asks the model to write a verification program for the task and to run this program on its output.
The model then revises the output where the verification fails.
\withappendix{Appendix~\ref{app:setup-details} describes how we apply them to the coding agent.}

\paragraph{Agent frameworks.}
We run \satisfix{} in two agent frameworks, OpenHands \citep{wang2025openhands} and mini-swe-agent \citep{yang2024sweagent}.
The latter gives the coding agent only a bash tool.
The base agent is the same framework without any modification.

\subsection{Comparison with Baselines}
\label{sec:baselines}

Table~\ref{tab:baselines} compares \satisfix{} with the three baselines and the base agent on DeepSWE with GPT-5.6-Luna.
\satisfix{} reaches a pass@1 of 50.22\%, averaged over four runs\withappendix{ (Appendix~\ref{app:passk} reports pass@$k$)}.
It outperforms the base agent by 5.97\% and Self-Refine by 3.98\%.

The three baselines improve pass@1 over the base agent by at most 1.99\%.
Self-Refine and ProgCo each reach 46.24\%, while DeCRIM stays at the 44.25\% of the base agent.
This result is consistent with the limitation in Section~\ref{sec:problem}.
An agent that has accepted $a^{\circ}$ can accept it again when it reviews its own work.

Self-Refine takes 70.2 steps per task (vs.\ 49.0 for the base agent) and uses about three times as many input tokens (10.14M vs.\ 3.37M).
\satisfix{} takes 58.6 steps per task and costs \$0.897, less than half as much as Self-Refine.

\begin{table}[t]
\centering
\small
\setlength{\tabcolsep}{3pt}
\begin{tabular*}{\columnwidth}{@{\extracolsep{\fill}}lccccc@{}}
\toprule
 & & & \multicolumn{2}{c}{\textbf{Tokens per task}} & \\
\cmidrule(lr){4-5}
\textbf{Method} & \textbf{Pass@1} & \textbf{Steps} & \textbf{Input} & \textbf{Output} & \textbf{Cost (\$)} \\
\midrule
Base & 44.25$_{\pm 2.98}$ & 49.0 & 3.37M & 25.8k & 0.644 \\
Self-Refine & 46.24$_{\pm 3.98}$ & 70.2 & 10.14M & 92.0k & 2.022 \\
DeCRIM & 44.25$_{\pm 2.61}$ & 47.9 & 3.39M & 25.6k & 0.645 \\
ProgCo & 46.24$_{\pm 1.82}$ & 51.1 & 4.12M & 35.3k & 0.809 \\
\satisfix{} & \textbf{50.22}$_{\pm 2.64}$ & 58.6 & 4.80M & 33.5k & 0.897 \\
\bottomrule
\end{tabular*}
\caption{\satisfix{}, the baselines, and the base agent (Base) on DeepSWE with GPT-5.6-Luna, four runs each. Pass@1 (\%) is the mean over runs $\pm$ the standard deviation. Steps, token counts, and cost (\$) are means per task.}
\label{tab:baselines}
\end{table}

\subsection{Cross-Model and Cross-Benchmark Comparison}
\label{sec:cross-model}

\begin{table*}[t]
\centering
\small
\setlength{\tabcolsep}{5pt}
\begin{tabular}{@{}l cc @{\hspace{14pt}} cc @{\hspace{10pt}} cc@{}}
\toprule
 & \multicolumn{2}{c}{\textbf{DeepSWE v1.1}} & \multicolumn{4}{c}{\textbf{SWE-bench Pro v2}} \\[3pt]
 & \multicolumn{2}{c}{Full (113)} & \multicolumn{2}{c}{Hard (51)} & \multicolumn{2}{c}{Full (642)} \\
\cmidrule(r{8pt}){2-3} \cmidrule(r{6pt}){4-5} \cmidrule(l){6-7}
\textbf{Model} & Base & \satisfix{} & Base & \satisfix{} & Base & \satisfix{} \\
\midrule
GPT-5.6-Luna      & 44.25$_{\pm 2.98}$ & \textbf{50.22}$_{\pm 2.64}$ & 78.43 & \textbf{84.31} & 91.12 & 92.37 \\
DeepSeek V4 Flash & 53.32$_{\pm 3.64}$ & \textbf{57.08}$_{\pm 1.69}$ & 78.43 & \textbf{82.35} & 92.21 & 92.37 \\
Qwen3.8-27B       & 42.26$_{\pm 2.33}$ & \textbf{47.79}$_{\pm 2.28}$ & 76.47 & \textbf{84.31} & 91.28 & 92.21 \\
Gemma 4 26B A4B   & -- & -- & 9.80 & \textbf{21.57} & 52.65 & \textbf{55.92} \\
Muse Glimmer 30B  & -- & -- & 37.25 & \textbf{43.14} & 80.37 & \textbf{83.02} \\
\bottomrule
\end{tabular}
\caption{Pass@1 (\%) of \satisfix{} and Base on DeepSWE v1.1 and SWE-bench Pro v2. On DeepSWE, each method runs four times, and pass@1 is the mean over runs $\pm$ the standard deviation. A dash (--) marks a model that hits a capability floor on DeepSWE.}
\label{tab:models}
\end{table*}

Table~\ref{tab:models} shows that the gain holds across models and benchmarks.
On DeepSWE with DeepSeek V4 Flash, \satisfix{} raises pass@1 from 53.32\% to 57.08\%.
It also saves 14.0\% of the token cost per task, as the coding agent takes fewer steps (131.6 vs.\ 152.9).
With Qwen3.8-27B, \satisfix{} raises pass@1 from 42.26\% to 47.79\%.
Gemma 4 26B A4B and Muse Glimmer 30B hit a capability floor on DeepSWE.

On the hard subset of SWE-bench Pro, \satisfix{} improves pass@1 with all five models by up to 11.77\%.
The largest gain is with Gemma 4 26B A4B, from 9.80\% to 21.57\%.
With GPT-5.6-Luna and Qwen3.8-27B, \satisfix{} gains 5.88\% and 7.84\% and reaches 84.31\% with both.

On the full set, \satisfix{} also improves pass@1 with all five models.
GPT-5.6-Luna, DeepSeek V4 Flash, and Qwen3.8-27B saturate the tasks of SWE-bench Pro v2 outside the hard subset.
They solve more than 92\% of these tasks with either method.
For these models, most of the room for improvement is in the hard subset.
With Gemma 4 26B A4B and Muse Glimmer 30B, \satisfix{} gains 3.27\% and 2.65\% on the full set.

\subsection{Cross-Framework Comparison}
\label{sec:cross-harness}

Table~\ref{tab:harnesses} shows that the gain also holds on OpenHands with GPT-5.6-Luna.
The base agent on OpenHands takes more than twice as many steps per task as on mini-swe-agent (118.8 vs.\ 49.0).
Its pass@1 is 46.90\%, against 44.25\% on mini-swe-agent.
\satisfix{} raises it to 51.33\%.
This gain of 4.43\% is close to the 5.97\% on mini-swe-agent.

\begin{table}[t]
\centering
\small
\setlength{\tabcolsep}{3pt}
\begin{tabular*}{\columnwidth}{@{\extracolsep{\fill}}lccccc@{}}
\toprule
 & & & \multicolumn{2}{c}{\textbf{Tokens per task}} & \\
\cmidrule(lr){4-5}
\textbf{Method} & \textbf{Pass@1} & \textbf{Steps} & \textbf{Input} & \textbf{Output} & \textbf{Cost (\$)} \\
\midrule
\multicolumn{6}{@{}l}{\textit{mini-swe-agent}} \\
Base & 44.25$_{\pm 2.98}$ & 49.0 & 3.37M & 25.8k & 0.644 \\
\satisfix{} & \textbf{50.22}$_{\pm 2.64}$ & 58.6 & 4.80M & 33.5k & 0.897 \\
\midrule
\multicolumn{6}{@{}l}{\textit{OpenHands}} \\
Base & 46.90$_{\pm 4.02}$ & 118.8 & 7.72M & 47.4k & 1.404 \\
\satisfix{} & \textbf{51.33}$_{\pm 2.40}$ & 127.4 & 9.42M & 58.0k & 1.714 \\
\bottomrule
\end{tabular*}
\caption{\satisfix{} and Base on two agent frameworks with GPT-5.6-Luna, four runs each. Columns as in Table~\ref{tab:baselines}.}
\label{tab:harnesses}
\end{table}

\subsection{Ablation Study}
\label{sec:ablation}

We ablate each part of \satisfix{} with GPT-5.6-Luna (Table~\ref{tab:ablation}).
\begin{itemize}[nosep,leftmargin=*]
  \item \textbf{w/o objective-aligned elicitation} removes the advisor role. It asks the coding agent directly whether it concluded that a requirement could be handled differently from what the task states.
  \item \textbf{w/o reconsideration} keeps the elicitation. After the report, \satisfix{} only puts the agent back in the developer's role. The agent then completes the task.
  \item \textbf{Self-correction in place of both} asks the agent at the same point to review its plan and check whether it has made a mistake. Like the first reconsideration message, it repeats the task description. The agent then completes the task.
\end{itemize}

Without objective-aligned elicitation, the pass rate falls from 49.6\% to 43.4\%, the level of the base agent.
Without reconsideration, it falls to 47.8\%.
Replacing both with self-correction gives 44.2\%, close to the base agent.
Objective-aligned elicitation accounts for most of the gain.

Without the advisor role, the agent does not have anything to report about a choice it has accepted (Section~\ref{sec:elicitation}).
Self-correction repeats the task description as the reconsideration message does, so rereading the task alone does not explain the gain of \satisfix{}.
Without reconsideration, part of the gain remains.
The report alone can bring the choices back to the agent's attention.

\begin{table}[t]
\centering
\small
\begin{tabular*}{\columnwidth}{@{\extracolsep{\fill}}lc@{}}
\toprule
\textbf{Variant} & \textbf{Pass rate (\%)} \\
\midrule
\satisfix{} (first run) & \textbf{49.6} \\
\quad w/o objective-aligned elicitation & 43.4 \\
\quad w/o reconsideration & 47.8 \\
\quad self-correction in place of both & 44.2 \\
Base (first run) & 43.4 \\
\bottomrule
\end{tabular*}
\caption{Ablation with GPT-5.6-Luna. \satisfix{} and Base are from their first runs.}
\label{tab:ablation}
\end{table}

\section{Analysis}
\label{sec:analysis}

We analyze the elicitation reports and reconsideration replies of all 113 DeepSWE tasks.
They come from the first run of \satisfix{} with GPT-5.6-Luna and DeepSeek V4 Flash.
This run has a pass rate of 49.6\% with GPT-5.6-Luna and 59.3\% with DeepSeek V4 Flash.
Claude Code \citep{anthropic2026claudecode} with Claude Opus 5 and Codex \citep{openai2026codex} with GPT-5.6-Sol annotated each task independently.
They split each elicitation report into items.
They labeled each item that reports specification satisficing with one of the four types of reasons in Table~\ref{tab:reasons}.
They also marked whether the coding agent reconsiders each such item in its reconsideration reply.
One of the authors checked every case where they disagreed and made the final decision.

\begin{figure}[t]
\centering
\iflatexml
\begin{minipage}{0.48\columnwidth}\centering
{\small GPT-5.6-Luna}\\
\includegraphics[width=\linewidth]{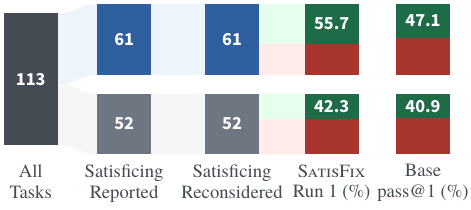}
\end{minipage}\hfill
\begin{minipage}{0.48\columnwidth}\centering
{\small DeepSeek V4 Flash}\\
\includegraphics[width=\linewidth]{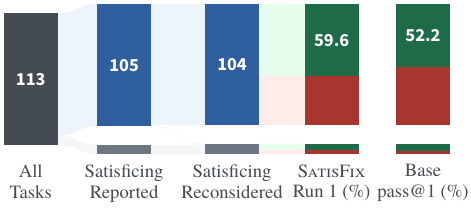}
\end{minipage}
\else
{\small GPT-5.6-Luna}\par\smallskip
\includegraphics[width=\columnwidth]{assets/fig-sankey-luna.pdf}\par\medskip
{\small DeepSeek V4 Flash}\par\smallskip
\includegraphics[width=\columnwidth]{assets/fig-sankey-deepseek.pdf}
\fi
\caption{The 113 tasks in the first run of \satisfix{}. In the second column, blue marks the tasks in which elicitation reports specification satisficing with one of the four types of reasons. In the third column, blue marks the tasks in which the agent reconsiders such an item. Gray marks the other tasks. The last two columns give the pass rate of \satisfix{} in this run and the pass@1 of the base agent over its four runs. Green is success and red is failure.}
\label{fig:sankey}
\end{figure}

\subsection{The Outcomes of Objective-Aligned Elicitation}
\label{sec:analysis-elicitation}

Objective-aligned elicitation brings out reports of specification satisficing in most tasks.
We count an item as specification satisficing when the agent considers a choice that meets the requirement and then leans toward one that does not.
Elicitation comes before implementation, so the agent sometimes reports that it remains unsure between the two.
We count these reports as well.
The report alone does not show whether the choice meets the requirement under $D$ (Definition~\ref{def:satisficing}).

With GPT-5.6-Luna, 90 items in 61 of the 113 tasks report specification satisficing with one of the four types of reasons in Table~\ref{tab:reasons}.
With DeepSeek V4 Flash, 227 items in 105 tasks do so.
Table~\ref{tab:ss-categories} shows that all four types occur with both models.
DeepSeek V4 Flash reports more than twice as many such items per task as GPT-5.6-Luna (2.01 vs.\ 0.80).

\begin{table}[t]
\centering
\small
\setlength{\tabcolsep}{2.5pt}
\begin{tabular*}{\columnwidth}{@{\extracolsep{\fill}}lrrrr@{}}
\toprule
 & \multicolumn{2}{c}{\textbf{Luna}} & \multicolumn{2}{c}{\textbf{DeepSeek}} \\
\cmidrule(lr){2-3}\cmidrule(lr){4-5}
\textbf{Type} & \textbf{Total} & \textbf{Avg.} & \textbf{Total} & \textbf{Avg.} \\
\midrule
Intended reinterpretation & 36 & 0.32 & 77 & 0.68 \\
Narrowed scope & 22 & 0.19 & 95 & 0.84 \\
Lowered standard & 19 & 0.17 & 32 & 0.28 \\
Perceived conflict & 13 & 0.12 & 23 & 0.20 \\
\midrule
\textbf{Total} & \textbf{90} & \textbf{0.80} & \textbf{227} & \textbf{2.01} \\
\bottomrule
\end{tabular*}
\caption{Items that report specification satisficing with GPT-5.6-Luna and DeepSeek V4 Flash, by the type of reason in Table~\ref{tab:reasons}. Avg.\ is the mean number per task over the 113 tasks.}
\label{tab:ss-categories}
\end{table}

\subsection{The Outcomes of Reconsideration}
\label{sec:analysis-reconsideration}

The coding agent reconsiders all 90 of these items with GPT-5.6-Luna and 226 of the 227 with DeepSeek V4 Flash.
Figure~\ref{fig:sankey} follows the tasks through both steps.

With GPT-5.6-Luna, the agent reconsiders such an item in 61 tasks.
On these tasks, \satisfix{} succeeds on 55.7\% and the base agent on 47.1\% over its four runs.
On the other 52 tasks, the two succeed on 42.3\% and 40.9\%.
With DeepSeek V4 Flash, \satisfix{} succeeds on 59.6\% of the 104 such tasks and the base agent on 52.2\%.
The other 9 tasks are too few to compare.

With GPT-5.6-Luna, \satisfix{} outperforms the base agent for every type of reason (Table~\ref{tab:reconsider}).
With DeepSeek V4 Flash, the gain for each type is at most 2.5\%, below the 7.4\% gain on all such tasks.
A task with several types counts under each of them.
With DeepSeek V4 Flash, \satisfix{} succeeds on 67.6\% of the 68 tasks with one or two such items and the base agent on 51.1\%.
On the 36 tasks with three or more, \satisfix{} succeeds on 44.4\% and the base agent on 54.2\%.
A possible reason is that the agent has built its plan on many such choices.
Reconsidering them all at once can then help less.
GPT-5.6-Luna rarely reports three or more.

\begin{table}[t]
\centering
\small
\setlength{\tabcolsep}{2pt}
\begin{tabular*}{\columnwidth}{@{\extracolsep{\fill}}lrrrrrr@{}}
\toprule
 & \multicolumn{3}{c}{\textbf{GPT-5.6-Luna}} & \multicolumn{3}{c}{\textbf{DeepSeek V4 Flash}} \\
\cmidrule(lr){2-4}\cmidrule(lr){5-7}
\textbf{Tasks} & \textbf{\#} & \textbf{Ours} & \textbf{Base} & \textbf{\#} & \textbf{Ours} & \textbf{Base} \\
\midrule
Any of the four types & 61 & 55.7 & 47.1 & 104 & 59.6 & 52.2 \\
\midrule
\multicolumn{7}{@{}l}{\textit{By type of reason}} \\
\quad Intended reinterp. & 32 & 53.1 & 49.2 & 55 & 56.4 & 54.5 \\
\quad Narrowed scope & 19 & 42.1 & 40.8 & 67 & 55.2 & 53.4 \\
\quad Lowered standard & 19 & 73.7 & 48.7 & 30 & 53.3 & 50.8 \\
\quad Perceived conflict & 12 & 50.0 & 39.6 & 21 & 52.4 & 52.4 \\
\midrule
\multicolumn{7}{@{}l}{\textit{By number of such items}} \\
\quad 1 & 34 & 58.8 & 51.5 & 39 & 69.2 & 53.8 \\
\quad 2 & 25 & 52.0 & 40.0 & 29 & 65.5 & 47.4 \\
\quad $\geq$3 & 2 & 50.0 & 62.5 & 36 & 44.4 & 54.2 \\
\bottomrule
\end{tabular*}
\caption{Results on the tasks in which the coding agent reconsiders an item with one of the four types of reasons in Table~\ref{tab:reasons}. Ours is the pass rate (\%) of \satisfix{} in its first run. Base is the pass@1 (\%) of the base agent over its four runs. A task with several types of reasons counts once for each type. \# is the number of tasks.}
\label{tab:reconsider}
\end{table}

\section{Related Work}
\label{sec:related}

\paragraph{Failure analysis of coding agents.}
Coding agents now handle long repository-level tasks \citep{yang2024sweagent,zhang2024autocoderover,wang2025openhands,huang2026deepswe}.
Benchmark studies report requirements that are omitted or only partly implemented \citep{zeng2025e2edevbench,jiang2026doc2feat,zhou2026featurebench}.
Failure analyses also report misinterpreted requirements \citep{liu2025failures,cemri2025mast}.
Other studies examine agent trajectories to find where a failure starts \citep{bouzenia2025trajectories,meng2026bugfixing,ceka2026traceability}.
\citet{wang2026solvedissues} show that some patches pass the tests and are still incorrect.
These studies classify failures by their outcome.
Our work studies failures in which the agent considers a choice that meets a requirement and later commits to one that does not.

\paragraph{Self-correction of LLMs and agents.}
Self-correction methods ask a model to find and fix the mistakes in its own output.
Self-Refine \citep{madaan2023selfrefine} and Reflexion \citep{shinn2023reflexion} ask the model to give feedback on its output and to revise it.
Another line of research adds a verification step.
Self-verification \citep{weng2023selfverification} and Chain-of-Verification \citep{dhuliawala2024cove} ask the model to verify its own answer.
DeCRIM \citep{palmeira-ferraz-etal-2024-llm} asks the model to check the output against each requirement.
CodeT \citep{chen2023codet} and ProgCo \citep{song-etal-2025-progco} check it with tests or programs that the model writes.
SCoRe \citep{kumar2025score} and RISE \citep{qu2024rise} train the model to correct itself over several turns.
However, LLMs rarely correct their own reasoning by themselves \citep{huang2024cannot,kamoi2024when,stechly2025self,tyen2024llms}.
Self-repair also brings limited gains in code generation \citep{olausson2024selfrepair}.
LLMs favor their own output when they review it \citep{xu-etal-2024-pride,panickssery2024llm}.
These methods rely on the model to judge whether its output contains a mistake.
In contrast, \satisfix{} asks the coding agent to report its implementation choices and their reasons without judging them.

\paragraph{Patch overfitting, specification gaming, and reward hacking.}
A system can satisfy the literal objective it is given and still miss what is intended.
In program repair, an overfitting patch passes the given tests without fixing the bug \citep{smith2015cure}.
In reinforcement learning, this is known as specification gaming or reward hacking \citep{krakovna2020specification,skalse2022defining,pan2022effects}.
LLMs exhibit it in context and when trained on human feedback \citep{pan2024feedback,wen2025mislead}.
Coding agents special-case or modify tests to make them pass \citep{zhong2026impossiblebench}.
In these cases, the system optimizes a signal that it can observe, such as a test result or a reward.
In contrast, a coding agent cannot observe whether its implementation choices are correct.
Specification satisficing is not overfitting to such a signal.
The agent accepts a choice that does not meet a requirement for a reason it finds acceptable.

\paragraph{Role prompting and elicitation.}
Assigning an LLM a role is widely used to steer its behavior and improve its reasoning \citep{kong2024roleplay}.
An LLM reports different information depending on the role it takes \citep{shanahan2023roleplay}.
Another line of research asks how to make a model report what it knows.
\citet{christiano2021elk} pose this problem as eliciting latent knowledge.
Auditing methods use personas and framings to draw out hidden objectives or secret knowledge \citep{marks2025auditing,cywinski2025secret}.
\citet{joglekar2025confessions} train a model to confess its own misbehavior.
These methods aim to improve answers or to audit a model.
In contrast, \satisfix{} puts the coding agent in the role of an advisor to the task's maintainers.
Reporting an implementation choice and its reason serves the objective of this role.
\satisfix{} then asks the agent to check each reported choice against the task description.

\section{Conclusion}
\label{sec:conclusion}

We presented \satisfix{}, an approach that helps a coding agent surface and mitigate specification satisficing.
\satisfix{} first asks the agent to report where it concluded that a requirement could be handled differently from what the task states.
It then asks the agent to check each reported choice against the task description.
\satisfix{} improves the pass@1 of the base agent by an average of 5.09\% on DeepSWE and 7.06\% on the hard subset of SWE-bench Pro.
On DeepSWE, it outperforms the best baseline by 3.98\%.

\label{body-end}

\section*{Limitations}

The elicitation message of \satisfix{} says that it comes from the task's maintainers.
We do not measure whether this changes how the coding agent treats later messages.
The Ethical Considerations section discusses this framing.

\satisfix{} relies on the agent's own answer to decide when to step in\withappendix{ (Appendix~\ref{app:trigger})}.
It also steps in only once per task, so it does not elicit choices that the agent commits to later.
We leave more reliable timing and repeated elicitation to future work.

Two LLM agents annotate the reports in Section~\ref{sec:analysis} independently.
An author checks only the items on which they disagree.
The author does not check items on which both agents make the same error.
The annotation also judges what the reports say, not whether the final code meets each requirement.

\section*{Ethical Considerations}

Prior work prompts LLMs with fictitious roles, stakes, or scenarios \citep{li2023emotionprompt,kong2024roleplay,marks2025auditing,cywinski2025secret}.
These prompts aim to improve or probe the behavior of LLMs.
\satisfix{} also presents its elicitation message as coming from the task's maintainers.
Unlike this work, it does not keep the framing.
Its next message tells the coding agent that the earlier message did not come from the maintainers.
It apologizes for presenting it that way and states that the task has not changed.

\bibliography{custom}

\end{document}